\documentclass[twoside]{article}
\usepackage{times}
\usepackage[latin1]{inputenc}
\usepackage{t1enc}
\usepackage{a4}
\usepackage{tabularx}
\usepackage{epsfig}
\usepackage{epsf,rotate}
\usepackage{graphicx}
\usepackage{epsf,epsfig,amsmath,amssymb}
\epsfclipon

\def\bref{\vspace{4pt}\noindent\hangindent=10mm}

\begin{document}

\pagestyle{myheadings}
\markboth{\hspace*{\fill}{\it \small
Norbert Christlieb
}}
{{\it \small
Finding the Most Metal-poor Stars
of the Galactic Halo
}\hspace{\fill}}

\vspace*{-1.4cm}
\thispagestyle{empty}

\noindent
{\scriptsize{\sc Astronomische Gesellschaft:} {\rm Reviews in
Modern Astronomy {\bf 16},
191--206 (2003)}}

\vspace*{2.0cm}

\begin{center}
{\Large\bf
Finding the Most Metal-poor Stars\\[0.0cm]
of the Galactic Halo with the\\[0.2cm]
Hamburg/ESO Objective-prism Survey}\\[0.7cm]

Norbert Christlieb\\[0.17cm]
Hamburger Sternwarte \\
Gojenbergsweg 112 \\
D-21029 Hamburg\\
Germany\\
E-Mail: nchristlieb@hs.uni-hamburg.de\\
URL: http://www.hs.uni-hamburg.de/DE/Ins/Per/Christlieb/
\end{center}

\vspace{0.5cm}

\begin{abstract}
{\it \noindent I review the status of the search for extremely
metal-poor halo stars with the Hamburg/ESO objective-prism survey
(HES). 2\,194 candidate metal-poor turn\-off stars and 6\,133
giants in the magnitude range $\mbox{\it 14} \lesssim B \lesssim
\mbox{\it 17.5}$ have been selected from 329 (out of 380) HES
fields, covering an effective area of 6\,400 square degrees in the
southern extragalactic sky. Moderate-resolution follow-up
observations for 3\,200 candidates have been obtained so far, and
$\sim \mbox{\it 200}$ new stars with $\mbox{\it [Fe/H]}<-\mbox{\it
3.0}$ have been found, which trebles the total number of such
extremely low-metallicity stars identified by all previous surveys.

We use VLT-UT2/UVES, Keck/HIRES, Subaru/HDS, TNG/SARG, and
Magellan/MIKE for high-resolution spectroscopy of HES metal-poor
stars. I provide an overview of the scientific aims of these
programs, and highlight several recent results. }
\end{abstract}

\section{Introduction}

Extremely metal-poor (EMP) stars, defined here as stars with
$1/1\,000$th of the metal content of the Sun (i.\,e.,
$\mbox{[Fe/H]}<-3.0$)\footnote{$\mbox{[A/B]} =
\log_{10} (N_{\mbox{\scriptsize A}}/N_{\mbox{\scriptsize B}}) -
\log_{10}(N_{\mbox{\scriptsize A}}/N_{\mbox{\scriptsize B}})_{\odot}$, for
elements A and B.}, permit astronomers to study the earliest epochs
of Galactic chemical evolution, since the atmospheres of these old,
low-mass stars retain, to a large extent, detailed information on
the chemical composition of the interstellar medium (ISM) at the
time and place of their birth. These stars are hence the local
equivalent of the high redshift Universe (for recent reviews on EMP
stars see Cayrel 1996, Beers 1999, 2000; Preston 2000).

In the past decade, numerous studies have focused on abundance
analysis of very metal-poor stars. In the early works of McWilliam
et\,al.\ (1995b) and Ryan et\,al.\ (1996), it was found that many
elements show a sudden change in the slope of their abundances
relative to Fe, near $\mbox{[Fe/H]}=-2.5$. Other elements,
including Ba and Sr, show a large scatter from star-to-star at very
low [Fe/H], indicating that the gas clouds from which these stars
formed were polluted by the products of only a few, if not single,
supernovae (see, e.\,g., Shigeyama \& Tsujimoto 1998; Tsujimoto
et\,al.\ 2000). Thus, with individual EMP stars we can indirectly
study the first generation of massive stars, which exploded in
supernovae of type~II (hereafter SN~II). Furthermore, Karlsson \&
Gustafsson (2001) have shown that the inspection of patterns seen
in abundance correlation diagrams of a large,
homogeneously-analyzed sample of stars allows one to constrain
theoretical yield calculations, and to perhaps even determine the
mass function of the first generation of stars.

EMP stars also contribute to observational cosmology: The
abundances of light elements (in particular Li), through models of
Big Bang nucleosynthesis, put limits on the baryonic content of the
Universe (e.\,g., Ryan et\,al.\ 1999), and individual age
determinations of the oldest stars -- either by nucleo-chronometry
(e.\,g., Cowan et\,al.\ 1999; Cayrel et\,al.\ 2001), or by means of
application of stellar evolutionary tracks (e.\,g., Fuhrmann 1998)
-- yield a lower limit for the age of the Galaxy, and hence the
Universe.

However, it should be recognized that, owing to their rarity, the
road to obtaining elemental abundances for EMP stars is long and
arduous. The process involves three major observational steps (see
Figure~1): (1) a wide-angle survey must be carried out, and
candidates for metal-poor stars have to be selected; (2)
moderate-resolution ($\sim 2$\,{\AA}) spectroscopic follow-up
observations of the candidates needs to be done in order to
validate the genuinely metal-poor stars among them; and (3)
high-resolution spectroscopy has to be obtained. Step (1) can be
carried in many different ways (for a review, see Beers 2000), for
example, with proper-motion surveys or photometric surveys.
However, spectroscopic surveys are the most efficient way to find
large numbers of EMP stars, and they have the advantage that the
samples they yield are not kinematically biased, as opposed to
proper-motion selected samples.

Until recently, the main source of EMP stars was the HK survey of
Beers and colleagues (Beers et\,al.\ 1985; Beers 1999), which so
far has yielded about $100$ stars with $\mbox{[Fe/H]}<-3.0$. For
most of them, high resolution ($R>40\,000$) spectroscopy has now
been obtained (see, e.\,g., Ryan et\,al.\ 1996; Norris et\,al.\
2001). As a consequence, the abundance trends of many elements are
now well-determined down to metalicities of $\mbox{[Fe/H]}\sim
-3.5$.

Still larger samples of EMP stars are needed to identify the small
subset of stars that are even more exotic then the EMP stars
themselves, e.\,g., stars with strong enhancements of the r-process
elements, suitable for nucleo-chronometric age dating (Cowan
et\,al.\ 1997, 1999; Cayrel et\,al.\ 2001; Hill et\,al.\ 2002). The
frequency of these stars among metal-poor giants with
$\mbox{[Fe/H]}<-2.5$ has been found in the HK survey to be as low
as $\sim 3$\,\%, a paucity confirmed by recent HES results (see
Section~4.2 for details). HE\,0107--5240 is another example of an
extremely rare star. It is a giant with $\mbox{[Fe/H]}=-5.3$
(Christlieb et\,al.\ 2002a), which was found during follow-up
observations of $\sim 200$ stars with $\mbox{[Fe/H]}<-3.0$ (see
Section~4.4 below).\linebreak

\hspace*{1cm}

\vspace*{-14mm}
\hspace*{-13mm}
\includegraphics[height=18cm,angle=0]{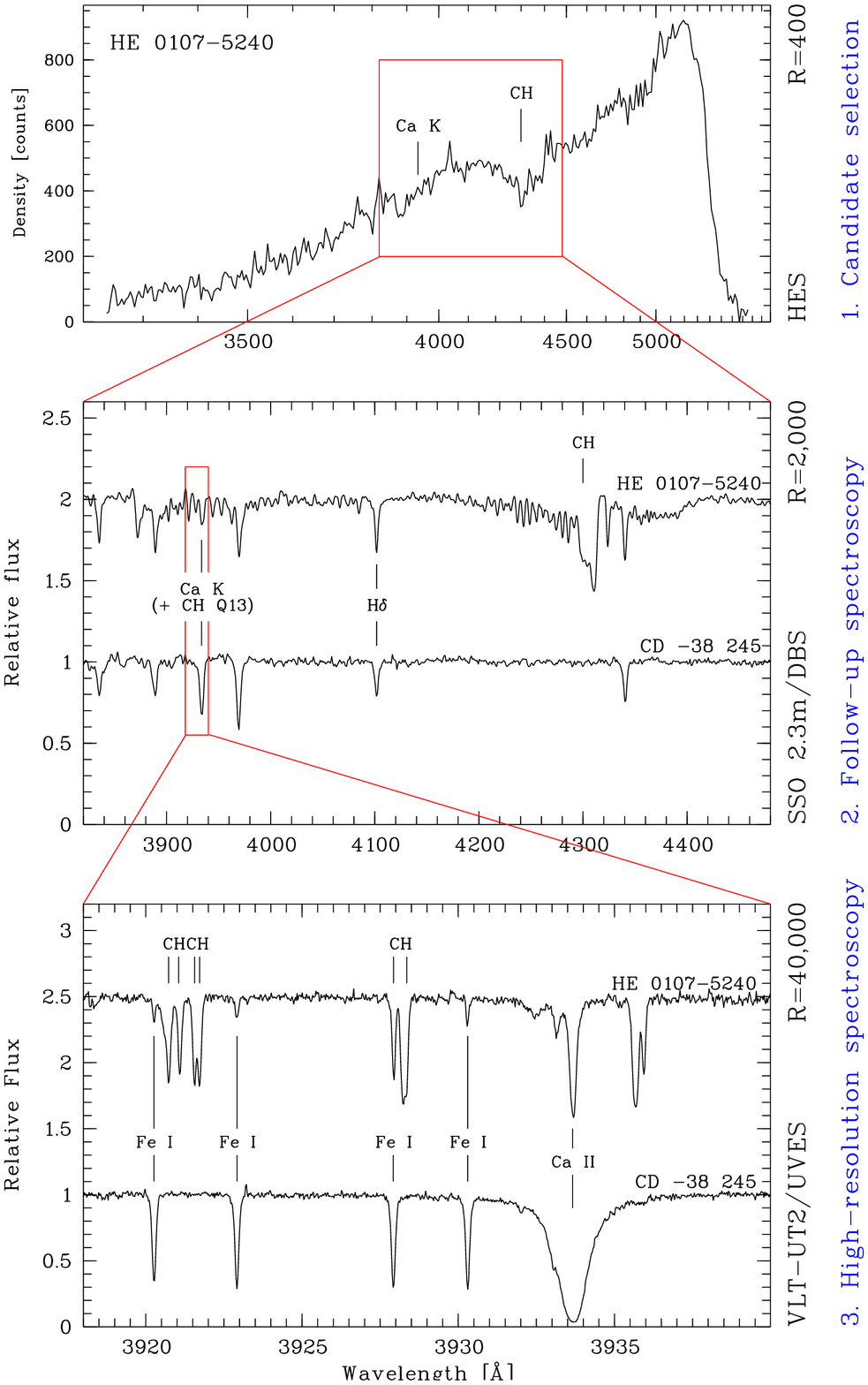}

\noindent
Figure 1: {\small The three major observational steps towards
obtaining elemental abundances of extremely metal-poor stars: (1)
objective-prism spectra, yielding candidate metal-poor stars; (2)
vetting of candidates by moderate-resolution follow-up
spectroscopy; (3) high-resolution spectroscopy of confirmed
candidates.}

\noindent
Furthermore, stars of the lowest metallicity place strong
constraints on the nature of the Metallicity Distribution Function
(MDF), which presumably arises from the ejected yields of the first
generations of stars. For example, with the exception of the star
at $\mbox{[Fe/H]} = -5.3$, it appears that the lowest abundance
stars in the Galaxy reach no lower than [Fe/H] $\sim -4.0$, hence
this is the level of early ``pollution'' of the ISM one might
demand to be reached by the products of first generation stars. One
also would like to test whether the MDF, as reflected by
observations of halo stars in the Solar neighborhood, remains
constant or varies with Galactocentric distance.

The Hamburg/ESO objective-prism survey (HES; Wisotzki et\,al. 2000)
offers the opportunity to increase the number of EMP stars by at
least a factor of five with respect to the HK survey, since the HES
is more than $1.5$\,mag deeper, and covers regions of the sky not
included in the HK survey (see Figure~2). For a detailed comparison
of the HES with the HK survey, see Christlieb \& Beers (2000).\\

\includegraphics{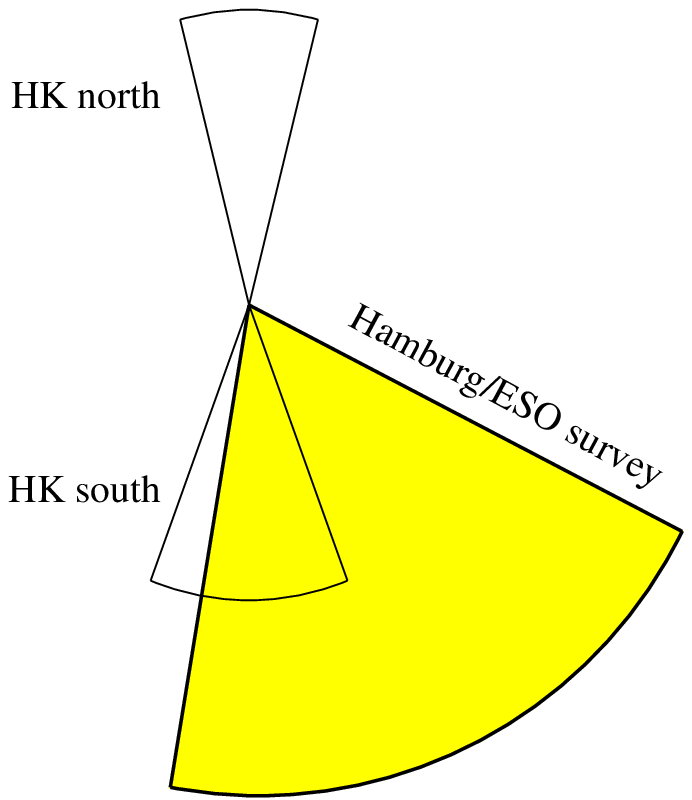}

\includegraphics{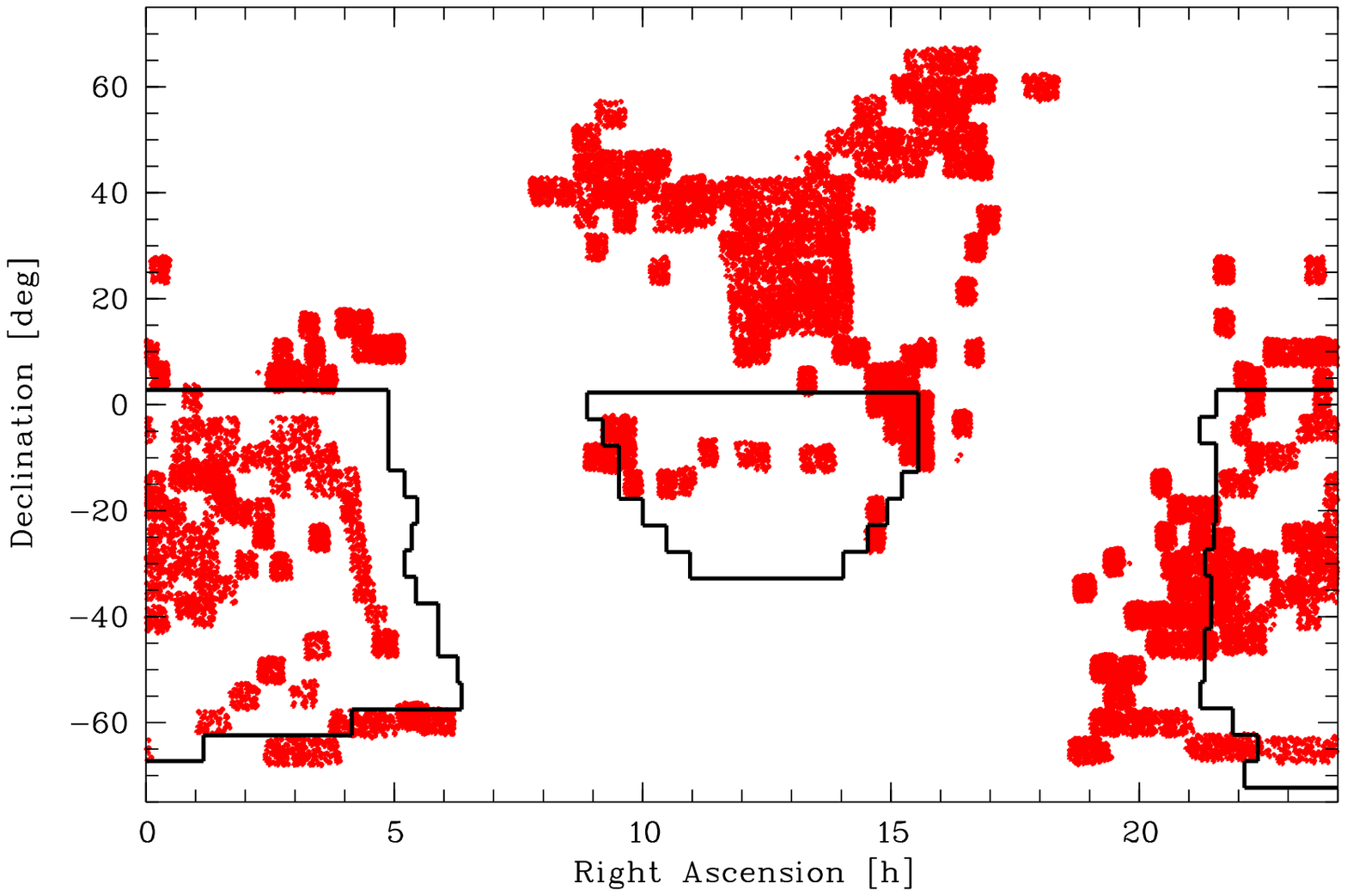}

\vspace*{55mm}
\noindent
Figure 2: {\small Comparison of the survey volumes of the HK
and HE surveys (left panel); comparison of HES area (framed) with HK
survey area (right panel). Grey areas denote HK survey plates.}\\

The HES covers the total southern ($\delta < -2.5^{\circ}$)
extragalactic ($|b| \gtrsim 30^{\circ}$) sky. The HES
objective-prism plates were obtained with the ESO 1\,m-Schmidt
telescope, and scanned and reduced at Hamburger Sternwarte. The
faintest HES metal-poor stars have $B\sim 17.5$\,mag; stars this
faint are observable at high spectral resolution using currently
existing telescope/instrument combinations.

\section{Selection of metal-poor candidates}

Metal-poor turnoff stars are selected in the HES database of
digital objective-prism spectra (for examples see Fig.~2) by
automatic spectral classification (Christlieb et\,al.\ 2002b).
Cooler stars ($B-V > 0.5$) are currently selected by applying an
empirical cutoff line in $(B-V)$ versus KP space, where KP is the
Ca {\sc II}~K line index defined by Beers et\,al.\ 1999).
\pagebreak

The selection has so far been restricted to unsaturated point
sources. For stars brighter than $B \sim 14$\,mag, saturation
effects (in particular in the red portions of the prism spectra)
begin to occur. Investigations on methods for recovering at least a
portion of these brighter candidates are underway (Frebel et\,al.,
in preparation). The importance of the bright sources comes from
the relative ease with which high-resolution spectroscopy can be
obtained. Extended sources are being rejected because spatial
extension leads to smeared-out spectral lines in slitless spectra,
and therefore many galaxies would otherwise enter the candidate
sample as false positives.

A total of $2\,194$ candidate metal-poor turnoff stars have been
selected from the 329 (out of 380) HES fields that are currently
used for stellar work. These fields cover a nominal area of
$8\,225\,\mbox{deg}^{2}$ of the southern extragalactic sky. The
covered area is reduced to an effective area of
$6\,400\,\mbox{deg}^{2}$ due to overlapping spectra (Wisotzki
et\,al.\ 2000; Christlieb et\,al.\ 2001a). In addition to the
turnoff stars, $6\,133$ metal-poor candidates with $0.5 < B-V <
1.2$ were selected, including $1\,785$ (i.e., $\sim 30$\,\%)
candidates expected to have large over-abundances of carbon
($\mbox{[C/Fe]}>1.0$). These stars are readily identifiable in the
HES from their strong CH G~bands (see Figure~4). A sample of 403
faint high latitude carbon (FHLC) stars has already been published
(Christlieb et\,al.\ 2001a), and that sample contains many
carbon-enhanced metal-poor (CEMP) stars as well, as recent
follow-up observations have demonstrated.

As noted above, the HES is a very rich source for metal-poor
giants. This is especially valuable because the present HK survey
sample is dominated by hotter stars near the main-sequence turnoff,
due to a temperature-related bias incurred by the initial visual
selection of HK candidates. As a result, only roughly 20\,\% of the
HK stars are metal-poor giants. However, efforts are under way to
recover metal-poor HK survey giants, by means of Artificial Neural
Network classifications of spectra extracted from digitized HK
survey plates (Rhee 2000) in combination with $JHK$ colors from the
2MASS survey (Skrutskie et\,al.\ 1997).

\section{Follow-up observations}

Moderate resolution ($\sim 2$\,{\AA}) follow-up observations are
needed to verify the metal-poor nature of the HES candidates. This
is the bottleneck in every large survey for EMP stars, and the only
way to cope with this is to establish large collaborations,
involving many people, with access to many different telescopes. To
provide the reader an impression of the extent of this effort, we
list in Table~1 the telescopes that have been used in 2002 for
follow-up of HES candidates, and the numbers of stars observed.

Follow-up spectra of $S/N > 20$ at Ca\,{\sc II}~K are required to
determine [Fe/H] of the metal-poor candidates with an accuracy of
0.2--0.3\,dex, using the technique of Beers et\,al.\ (1999). As can
be seen from Table~1, on average $\sim 20$ stars per night can be
observed at a 4\,m-class telescope in single-slit mode, taking into
account losses due to bad weather, moonlight, and technical
downtime. Note that the candidate throughput at the 1\,m-UK Schmidt
is as high achieved as by telescopes with $\sim 15$ times more
collecting area, due to the availability of the 150 fiber
instrument 6dF\linebreak

\hspace*{1cm}

\vspace*{-15mm}
\begin{center}
\epsfig{file=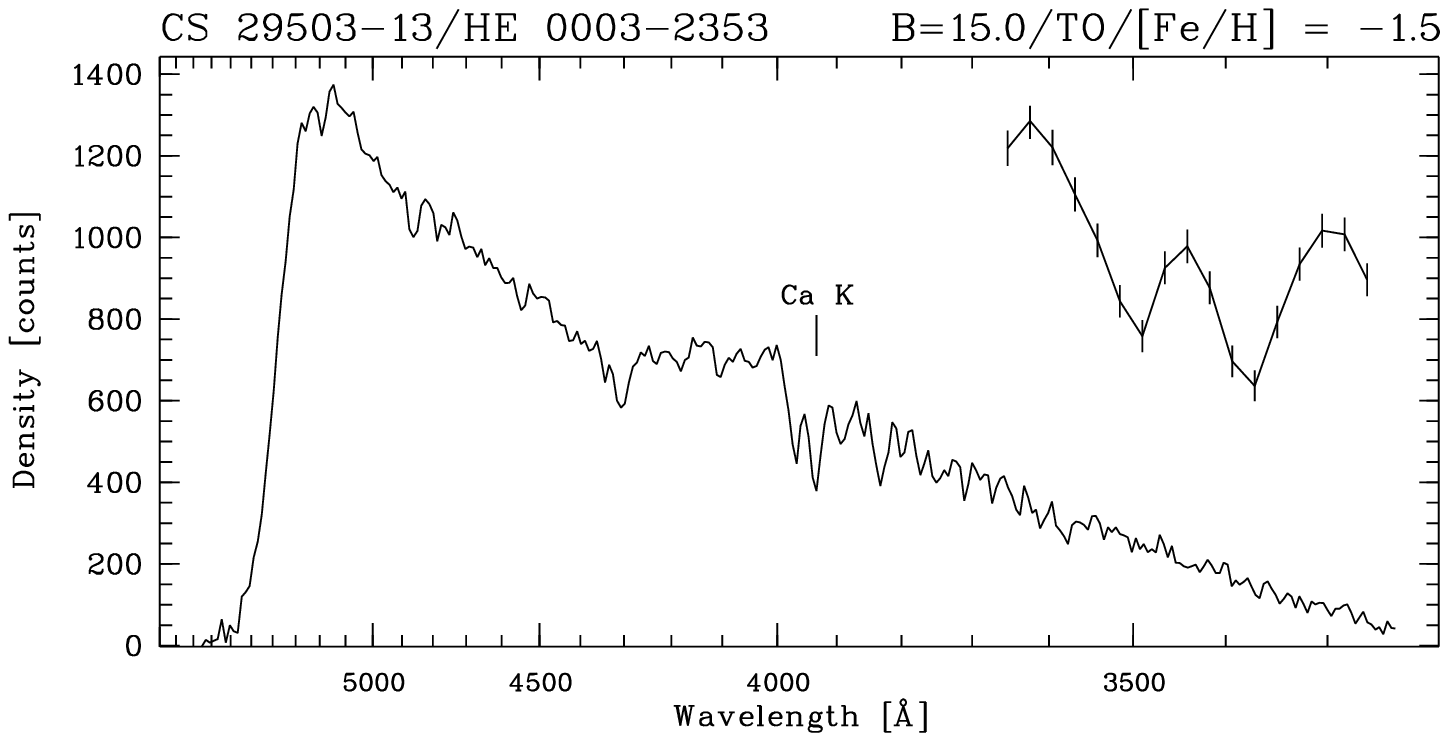, clip=, width=11.8cm,
      bbllx=53, bblly=452, bburx=470, bbury=638}\\[3mm]
\epsfig{file=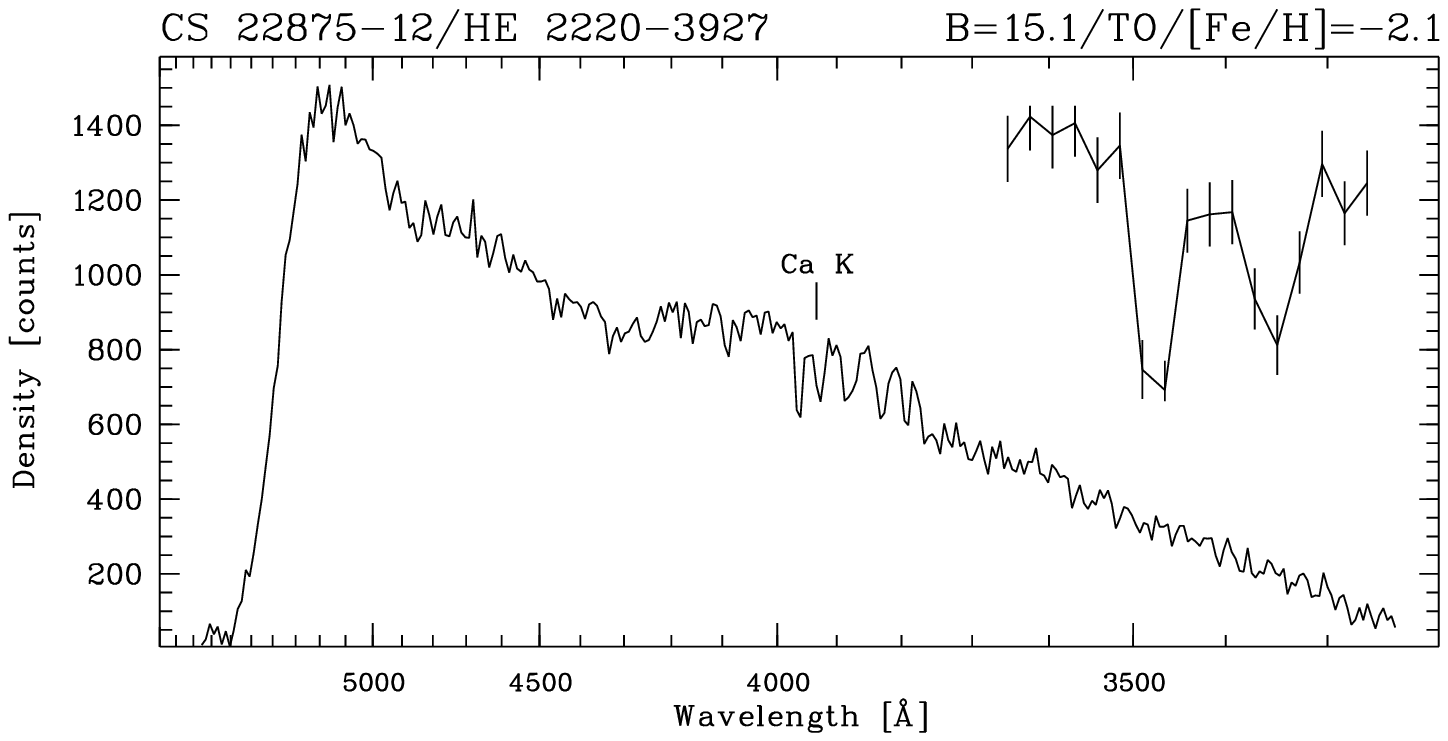, clip=, width=11.8cm,
      bbllx=53, bblly=452, bburx=470, bbury=638}\\[3mm]
\epsfig{file=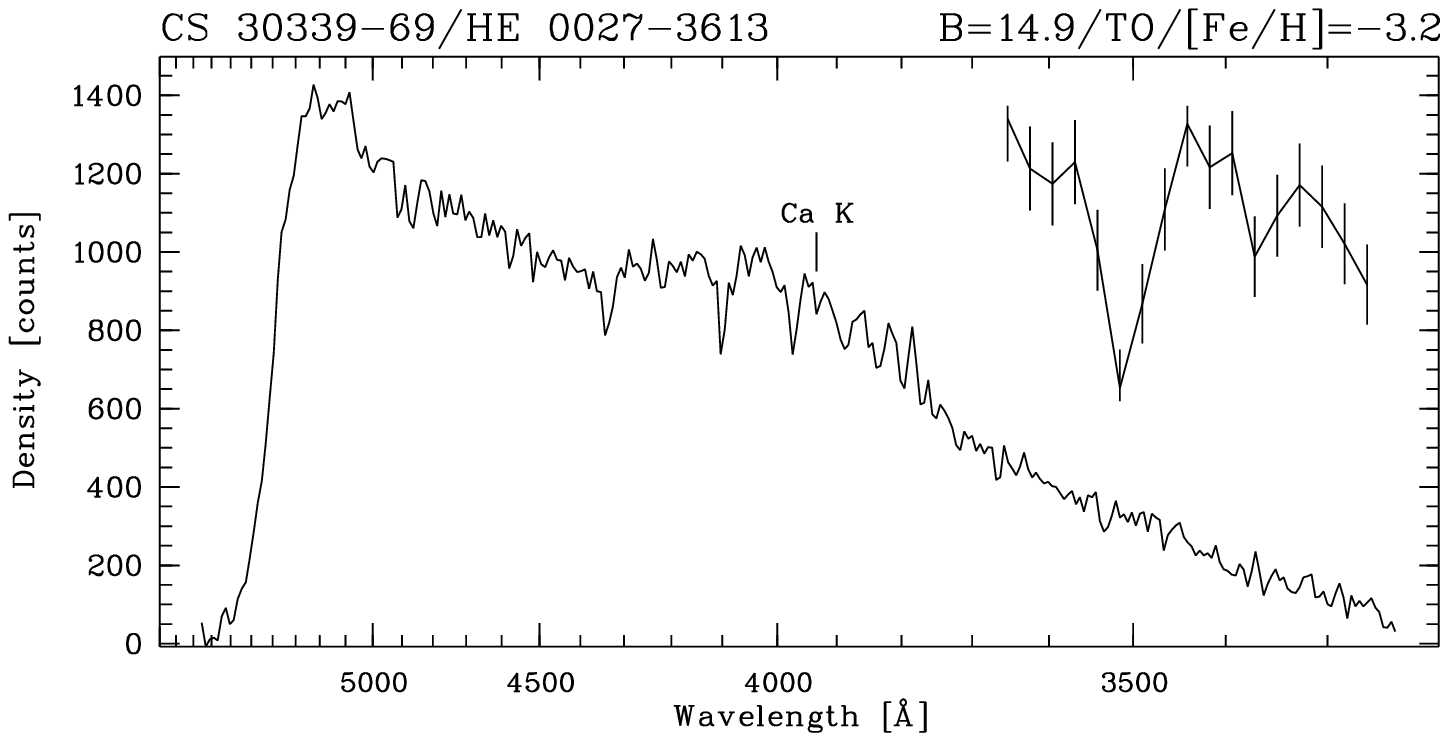, clip=, width=11.8cm,
      bbllx=53, bblly=426, bburx=470, bbury=638}
\end{center}

\noindent
Figure 3: {\small HES spectra of metal-poor stars found in the HK
survey. Note that wavelength is {\sl de}creasing from left to
right. The sharp cutoff at $\sim 5400$\,{\AA} is due to the IIIa-J
emulsion sensitivity cutoff (``red edge'').}
\vfill
\pagebreak

\hspace*{1cm}

\vspace*{-3mm}
\hspace*{-7mm}
\epsfig{file=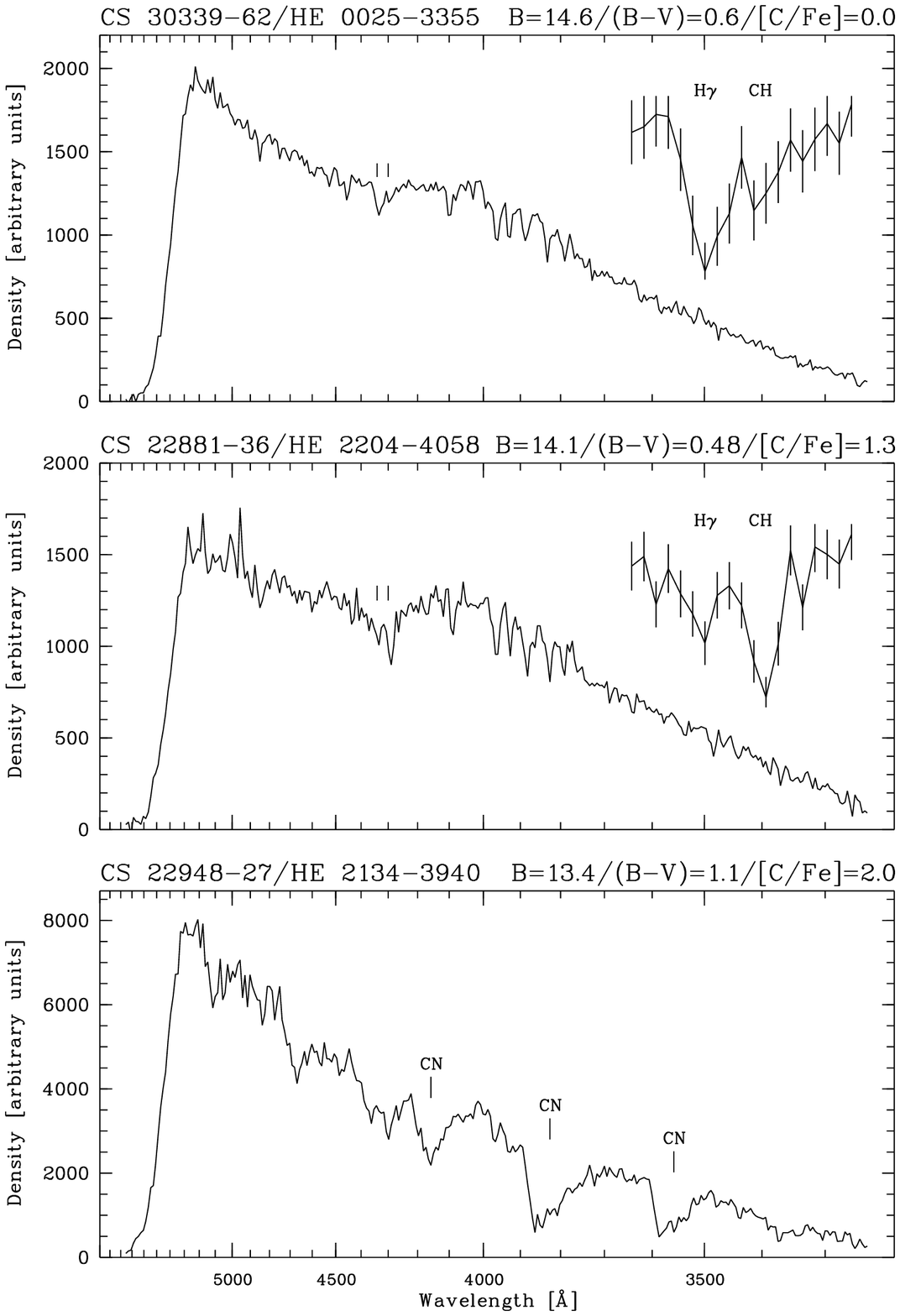, clip=, width=12.0cm,
    bbllx=50, bblly=170, bburx=475, bbury=781}

\begin{center}
Figure 4: {\small HES spectra of carbon-enhanced metal-poor stars
from the HK survey.}
\end{center}
\vfill
\pagebreak

\begin{center}
Table 1: Follow-up observations of HES metal-poor candidates in
2002.
\end{center}

\vspace*{-3mm}
\begin{table}[htbp]
  \begin{center}
    \begin{tabular}{lrrc}\hline\hline
 Telescope/Instr.   & Nights & Stars & Stars/Night \rule{0ex}{2.5ex}\\\hline
 AAT/RGOS           & $  6$ & $  87$ & $15$ \\
 ESO 3.6m/EFOSC2    & $ 13$ & $ 274$ & $21$ \\
 KPNO 4m/RCSP       & $  5$ & $ 109$ & $22$ \\
 Magellan~I/B\&C    & $ 33$ & $ 713$ & $22$ \\
 Palomar $200''$/DS & $  9$ & $  96$ & $11$ \\
 SSO 2.3\,m/DBS     & $ 32$ & $ 430$ & $13$ \\
 UK Schmidt/6dF     & $ 16$ & $ 341$ & $21$ \\\hline
 Sum/average        & $114$ & $2050$ & $18$ \rule{0ex}{2.5ex} \\\hline\hline
      \end{tabular}
    \end{center}
\end{table}

\noindent
(Watson 1998). Up to three fields with $10$--$60$ metal-poor
candidates, respectively, can be observed in one night. (The
remaining fibers are filled with other interesting HES stars,
e.\,g., field horizontal-branch stars, sdB stars, and white
dwarfs.) However, the faintest HES candidates ($B\gtrsim 16.5$)
cannot be observed with UK~Schmidt/6dF, since the fiber diameter is
$6.7''$ on the sky (chosen to suite the 6dF Galaxy Redshift
Survey), and therefore observations of point sources are
sky-background limited.

As of 1 February 2003, follow-up observations for $3\,200$ HES
candidates have been obtained. Considering that an average pace of
$\sim 1\,000$ candidates per year is the maximum one can hope to
reach in the long run, and the fact that $\sim 1\,000$ additional,
bright candidates are expected to be found in the HES, the
follow-up will likely continue for another 3--5 years.

The ``effective yield'' of metal-poor stars in the HES, as compared
to the HK survey, is presented in Table~2, and illustrated in
Figure~5. The effective yield (EY) refers to the fraction of
genuine stars below a given $\mbox{[Fe/H]}$ in the observed
candidate sample, as defined by Beers (2000). Based on a sample of
$1\,214$ candidates, it appears that the HES is twice as efficient
as the HK survey in finding stars with $\mbox{[Fe/H]}<-3.0$, when
photoelectric $BV$ photometry is used in the latter survey for
pre-selection, and six times as efficient compared to the HK survey
without usage of $BV$ photometry. The higher yields of the HES can
mainly be attributed to the fact that $B-V$ colors can be derived
directly from the objective-prism spectra with an accuracy of
better than $0.1$\,mag (Christlieb et\,al.\ 2001b), the higher
quality of the HES spectra, and the employment of quantitative
selection criteria, as opposed to visual selection in the HK
survey.

It should be stressed that a bias towards higher EYs may be present
in the HES candidate sample investigated so far, because the best
(in terms of the weakness of the Ca {\sc II}~K lines visible in the
HES spectra) and brightest candidates have been observed first.
However, when the EYs within each of the candidate classes (as
assigned during the visual verification of the HES candidates) are
applied to the respective fraction of candidates within the total
set of HES candidates, overall EYs of 4\,\% for stars with
$\mbox{[Fe/H]}<-3.0$ result for turnoff stars and giants. If more
restrictive selection criteria were to be used, the EYs could be
increased to $\sim 10$\,\%, but in this case $\sim 1/4$ of the
stars with $\mbox{[Fe/H]}<-3.0$ would not be identified.\linebreak

\noindent
Table 2: ``Effective yields'' of metal-poor stars (i.\,e., the
fraction of genuine metal-poor stars below a given [Fe/H] in the
observed candidate sample) in the HK and the HE surveys. $N$ refers
to the number of stars from which the statistic has been derived.

\begin{center}
    \begin{tabular}{lccrrr}\hline\hline
      \rule{0ex}{2.3ex} &          & \multicolumn{3}{c}{[Fe/H]}\\[-1mm]
 {\Large $^{\rm Survey}$} & {\Large $^{N}$} & $<-2.0$ & $<-2.5$ & $<-3.0$\\\hline
      HK survey/no   $B-V$\rule{0ex}{2.3ex} & 2614 & 11\,\% &  4\,\% & 1\,\%\\
      HK survey/with $B-V$\rule{0ex}{2.3ex} & 2140 & 32\,\% & 11\,\% & 3\,\%\\
      HES (turnoff stars)                   &  571 & 59\,\% & 21\,\% & 6\,\%\\
      HES (giants)                          &  643 & 50\,\% & 20\,\% & 6\,\%\\\hline\hline
    \end{tabular}
\end{center}

\vfill

\begin{center}
\epsfig{file=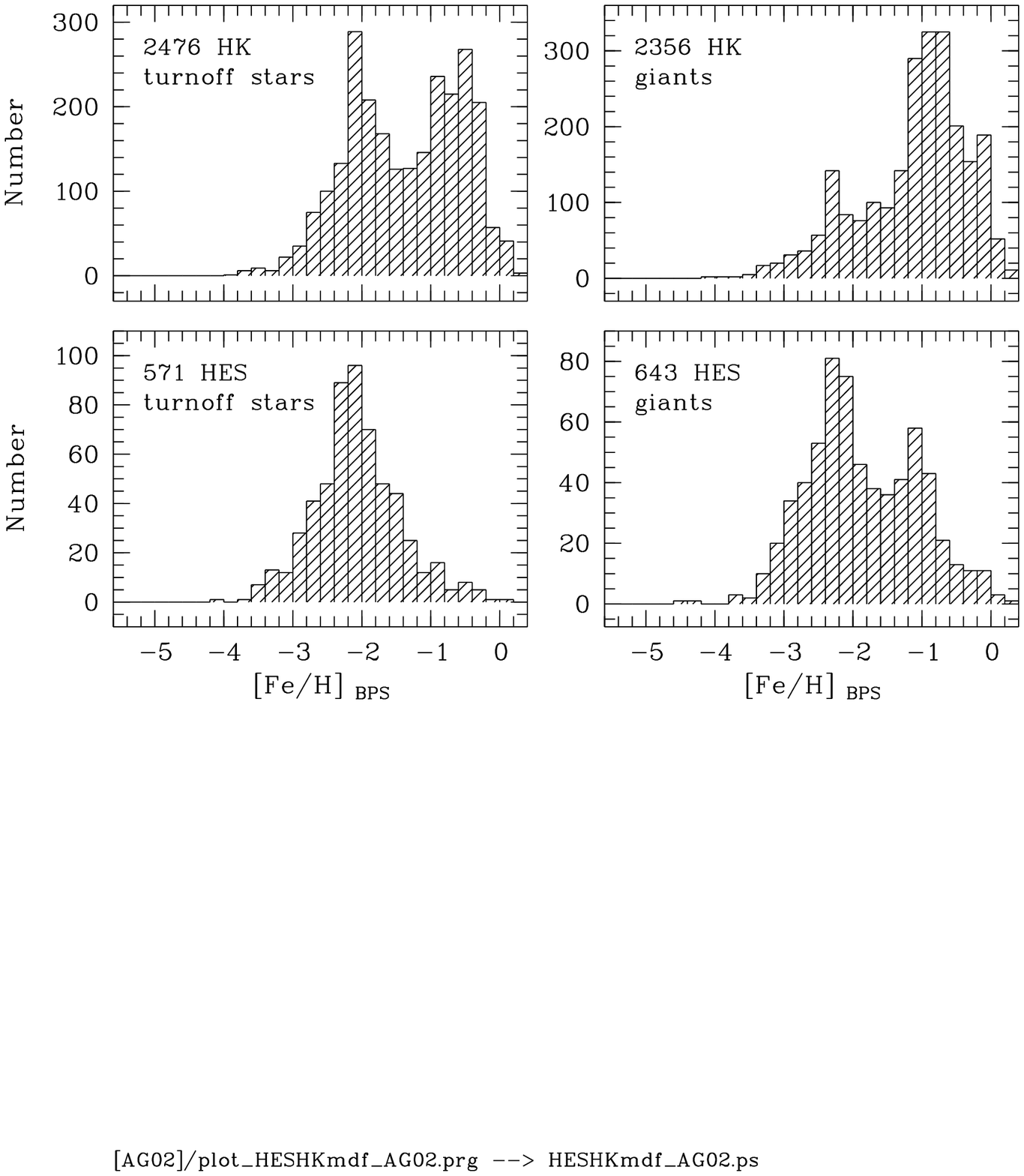, clip=, width=11.5cm,
      bbllx=31, bblly=346, bburx=521, bbury=685}
\end{center}

\noindent
Figure 5: {\small The observed Metallicity Distribution Functions
of HES and HK turnoff stars (left panel) and giants (right
panel)}

\vfill

\noindent
Hence, a compromise is struck between the ``purity'' of the sample
and the desire to include as many EMP stars as possible.

Using the numbers mentioned above, it can be estimated that at least $300$
stars with $\mbox{[Fe/H]}<-3.0$ will eventually be found in the HES, provided that
follow-up observations of \emph{all} candidates are obtained
(this estimate does not count the additional bright HES candidates, follow-up
observations for which have only recently begun).
\pagebreak

\section{High-resolution spectroscopy projects}

In this section, I provide an overview of ongoing high-resolution
spectroscopic analyses of confirmed metal-poor stars (mainly) from
the HES, and highlight some recent results.

\subsection{The 0Z (zero Z) project}

In a collaboration between Hamburger Sternwarte, Caltech, The
Observatories of the Carnegie Institution of Washington (and
formerly also Padua Observatory), we aim at obtaining
high-resolution, high $S/N$ spectroscopy of a large sample of the
most metal-poor stars, with Keck/HIRES and Magellan/MIKE. First
results have been published published in Cohen et\,al.\ (2002),
Carretta et\,al.\ (2002), Lucatello et\,al.\ (2003), and Cohen
et\,al.\ (2003).

Highlights include the discovery of HE\,0024--2523. It is a
peculiar main-sequence turnoff star, and a spectroscopic binary
with a period of 3.14 days. Its abundance pattern is dominated by
extreme enrichment of s-process elements, likely due to mass
accretion from an already evolved companion which has now probably
become a white dwarf. The relative lead abundance is
$\mbox{[Pb/Fe]}=+3.3$\,dex; the abundance ratio of Pb to Ba is even
more extreme, $\mbox{[Pb/Ba]}=+1.9$\,dex, among the highest ever
determined in a metal-poor star. The unusually short period of this
CH star suggests a past common envelope phase.

Stars like HE\,0024--2523 provide the unique opportunity to study
an extinct generation of extremely metal-poor AGB stars that left
their fingerprints on their less massive companions. These stars
place strong observational constraints on simulations of the
structure, evolution, and nucleosynthesis of AGB stars, and allow
us to study the s-process at very low metalicities. Recent
calculations of Goriely \& Siess (2001) predict that an efficient
production of s-process elements takes place even in zero
metalicity AGB stars, despite the absence of iron seeds, provided
that protons are mixed into carbon-rich layers. Proton mixing
results in the formation of $^{13}$C, which is a strong neutron
source due to the reaction $^{13}$C($\alpha$, n)$^{16}$O. The
strong over-abundances of Pb in HE\,0024--2523 and other ``Pb
stars'' (Aoki et\,al.\ 2000; VanEck et\,al.\ 2001; Aoki et\,al.\
2002b) are in concert with these predictions.

Another exotic star found with Keck/HIRES observations is the EMP
turnoff star HE~2148$-$1247. It appears that this star is enriched
in both its r- {\sl and} s-process elements, similar to
CS~22898-027 (Preston \& Sneden 2001; Aoki et\,al.\ 2002b), which
is almost a twin of HE~2148$-$1247 with respect to stellar
parameters as well as abundances, and a few other stars in the
literature (Preston \& Sneden 2001; Aoki et\,al.\ 2002b; Hill
et\,al.\ 2000).

Cohen et\,al.\ (2003) propose a scenario for explaining the
abundance pattern of HE~2148$-$1247. They suggest that the star is
a member of a binary system (and in fact, the star shows
significant radial variations) in which the former primary went
through the AGB phase. In this stage, carbon and s-process elements
were dumped onto the surface of HE~2148$-$1247 via mass transfer.
The former primary then evolved into a white dwarf. Mass transfer
then occurred into the {\it reverse} direction, i.\,e., from
HE~2148$-$1247 to the white dwarf, which led to an
accretion-induced collapse of the white dwarf into a neutron star.
R-process nucleosynthesis then occurs in a neutrino-driven wind of
the neutron star, and the nucleosynthesis products contaminate the
surface layers of the star that we see today.

\subsection{The HERES project -- An ESO Large Programme}

EMP stars with strong enhancements of r-process elements, like
CS~22892-052 (McWilliam et\,al.\ 1995; Sneden et\,al.\ 1996; Cowan
et\,al.\ 1997), and CS~31082-001 (Cayrel et\,al.\ 2001; Hill
et\,al.\ 2002), are suitable for individual age determinations
through the detection of the long-lived radioactive species thorium
and/or uranium. Such old stars can therefore provide lower limits
to the age of the Galaxy, and hence the Universe.

However, it appears that the pairs of elements used as chronometers
need to be chosen very carefully. The results of Hill et\,al.\
(2002) suggest that comparison of the abundance of Th with that of
the stable r-process element Eu does not yield reliable ages for
all stars, and in particular an unrealistically small age for
CS~31082-001. Th/Eu is roughly a factor of three higher in
CS~31081-001 than observed in CS~22892-052, yielding a
radioactive decay age for CS~31082-001 that is younger than the
Sun! U/Th is expected to be a more accurate chronometer, due to the
fact that $^{238}$U, with a half-life as ``short'' as 4.5\,Gyr (for
comparison, $^{232}$Th has a half-life of 14\,Gyr), is involved.
Furthermore, recent theoretical studies on r-process
nucleosynthesis (Wanajo et\,al.\ 2002; Schatz et\,al.\ 2002) reveal
that this chronometer is less sensitive to uncertainties of nuclear
physics than any other chronometer that is amenable to observation
in EMP stars.

Unfortunately, it is observationally very challenging to detect
uranium in metal-poor stars; CS~31081-001 was the first one for
which such a detection has been made. This is because the strongest
spectral line in the optical, U~II $3859.57$\,{\AA}, is very weak
(see Figure 9 of Hill et\,al.\ 2002), even if the enhancement of
r-process elements is as high as $\mbox{[r/Fe]}=1.7$\,dex, as
observed in CS~31081-001. Therefore, spectra of $S/N>200$ per pixel
and $R\ge 75\,000$ are required for derivation of accurate U
abundances. Also, one has to contend with the difficulty that, if
the carbon abundance is significantly enhanced with respect to
iron, as is the case for CS~22892-052 ($\mbox{[C/Fe]}=0.98$;
McWilliam et\,al.\ 1995), the U line is blended with a CN line. As
a result, measurement of an U abundance is impossible
for\linebreak
CS~22892-052.

Quite apart from their obvious use in individual age determination,
strongly r-process-enhanced metal-poor stars allow one to study
r-process nucleosynthesis in general, and possibly to identify the
site(s) of the r-process. While the abundance pattern of
CS~22892-052 for the heavier stable neutron-capture elements
($\mbox{Z}\ge 56$), and the abundances of the elements in the range
$56\le\mbox{Z}\le 72$ in CS~31082-001 match a scaled solar
r-process pattern very well, deviations are seen for some of the
lighter neutron-capture elements in both stars, and for Th (as
discussed above) and possibly Os in CS~31082-001. It has therefore
been suggested (Sneden et\,al.\ 2000; Hill et\,al.\ 2002) that (at
least) two r-processes must be at work.

Given these results, it is clear that a much larger sample of
strongly r-process- enhanced metal-poor stars is required. The {\sl
H}\/amburg/{\sl E}\/SO {\sl R}\/-processed {\sl E}\/nhanced star
{\sl S}\/urvey is aiming to increase the number of strongly
r-process-enhanced stars to (conservatively) $\sim 5$--10, and to
study a major fraction of these by means of high-resolution, high
signal-to-noise UVES spectra. The project is carried out in the
framework of an ESO Large Programme (P.\,I. Christlieb), which was
recently approved for an initial period of one year.

Stars with strong r-process enhancement can be readily recognized
from their prominent Europium lines. For example, CS~22892-052 has
an Eu~II $4129.7$\,{\AA} line as strong as $132$\,m{\AA}. This
means that r-process-enhanced stars can be identified from
``snapshot'' high-resolution spectra, i.e., $R=20\,000$ and
$S/N=30$. Such spectra can easily be obtained, e.\,g., with
VLT-UT2/UVES, or with Subaru/HDS, in 20\,min (including overheads)
for a $B=15$\,mag star, even under unfavorable conditions (e.\,g.,
full moon, seeing $2''$, thin clouds). About half of our follow-up
efforts presented above are seeking confirmation of giants with
[Fe/H] $\le -2.5$ for this project.

To date, a total of 152 targets, 128 of them from the HES, have
been observed in snapshot mode with UVES. Among the 98 stars with
$\mbox{[Fe/H]}< -2.5$\,dex, four with $\mbox{[r/Fe]}>+1.0$\,dex
have been found (three of them are HES stars). Among the 54 stars
with $\mbox{[Fe/H]}\ge -2.5$\,dex, one with $\mbox{[r/Fe]}>
+1.0$\,dex has been found. Higher resolution and $S/N$ spectra of
most of these stars is currently being obtained with VLT/UVES, for
attempts to detect uranium. An additional set of 117 stars is
scheduled to be observed in snapshot mode in Period 72
(April--September 2003).

\subsection{Carbon-enhanced metal-poor stars}

One of the early results from the HK survey was the finding that
the frequency of carbon-enhanced stars increases dramatically at
the lowest values of [Fe/H], reaching $\sim 20$--25\,\% at
$\mbox{[Fe/H]}<-2.5$ (Norris et\,al.\ 1997a; Rossi et\,al.\ 1999).
This result has been confirmed by follow-up observations of HES
candidates.

It turns out that nature presents us with a ``zoo'' of
carbon-enhanced metal-poor (CEMP) stars, with vastly differing
properties. Above all, the origin of carbon in this class of stars
remains unclear, despite extensive investigations by means of
high-resolution spectroscopy (e.\,g., Norris et\,al.\ 1997a,b;
Bonifacio et\,al.\ 1998; Hill et\,al.\ 2000; Aoki et\,al.\ 2002a;
Norris et\,al.\ 2002; Lucatello et\,al.\ 2003). A larger sample of
CEMP stars with abundance determinations, as well as long-term
radial velocity monitoring, is needed to constrain the possible
carbon enrichment scenarios realized in nature, and to separate
known CEMP stars into the corresponding classes of objects.

We are currently increasing the number of CEMP stars in dedicated
spectroscopic follow-up observations, mainly at the CTIO and KPNO
4\,m telescopes. Confirmed CEMP stars from this and other efforts
are studied at high resolution with Subaru/HDS and TNG/SARG. A
program aiming at radial velocity monitoring of a large sample of
CEMP stars has been proposed to be carried out with the
Anglo-Australian Telescope, and will hopefully be continued for
many years with this, and other 4m-class telescopes in the future.
\pagebreak

\vspace*{-8mm}
\begin{center}
\epsfig{file=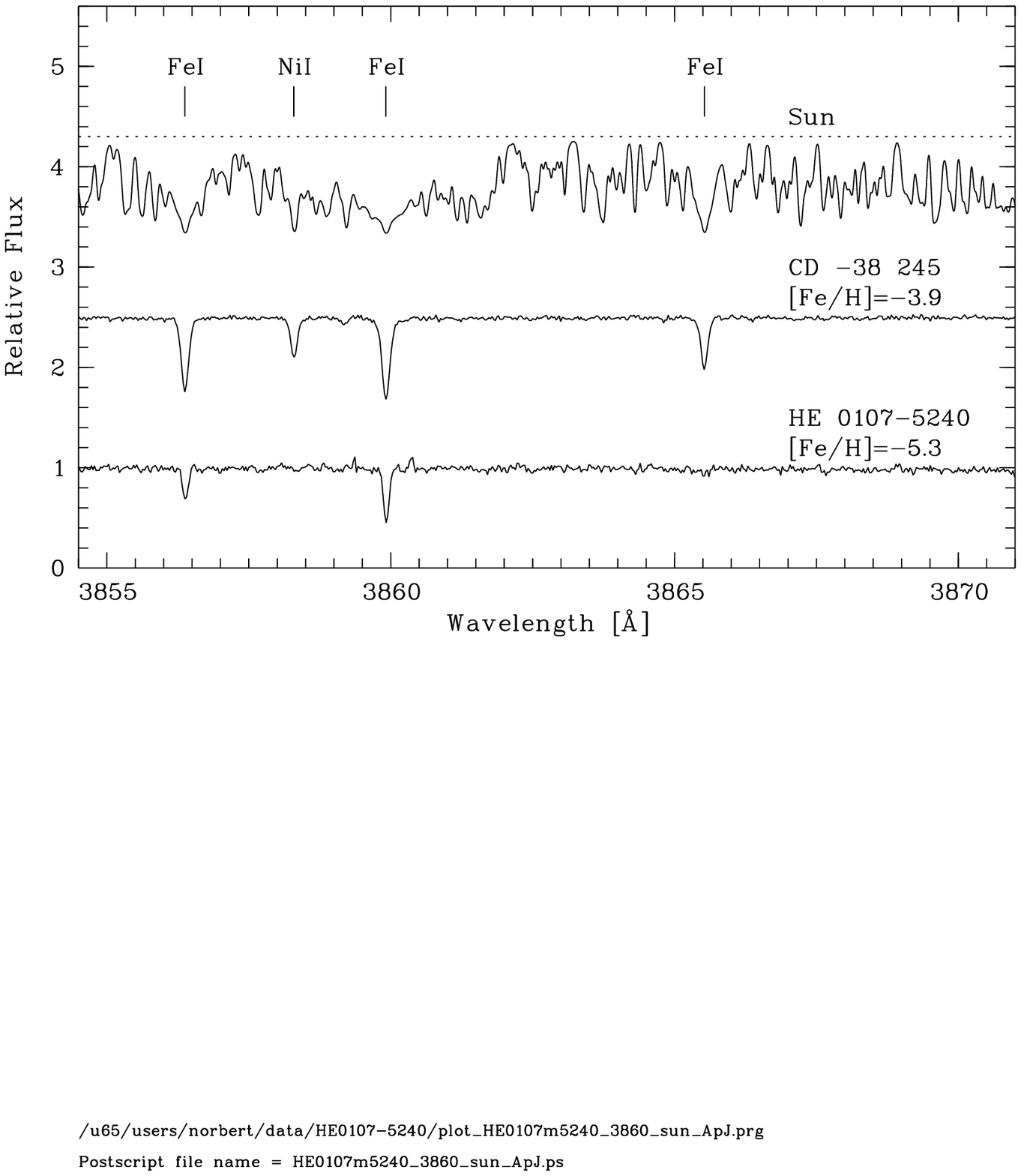, clip=, width=11.8cm,
  bbllx=62, bblly=335, bburx=527, bbury=627}\\[2ex]
\epsfig{file=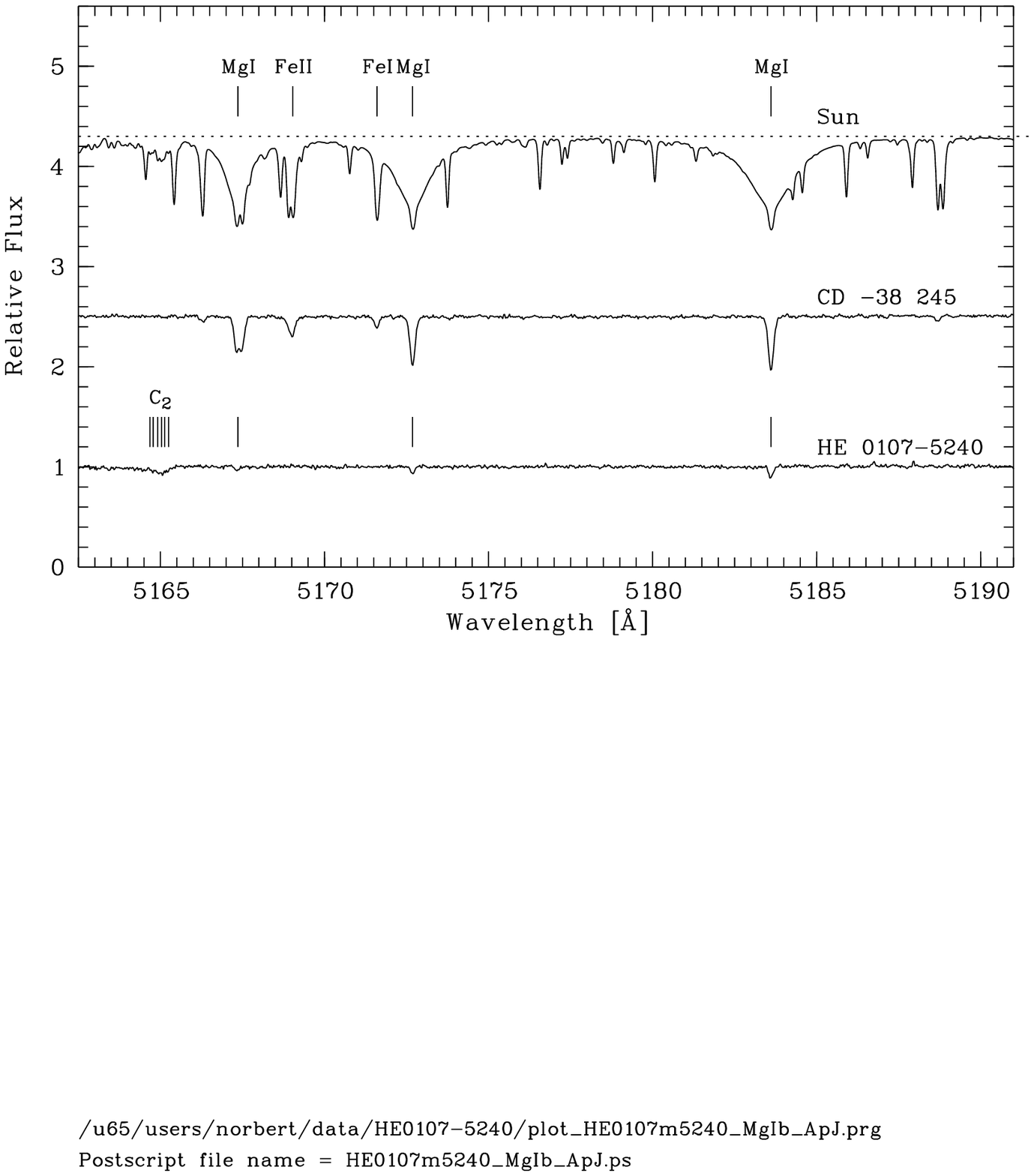, clip=, width=11.8cm,
  bbllx=62, bblly=335, bburx=527, bbury=627}    
\end{center}

\noindent
Figure 6: {\small Spectrum of the Sun compared with the VLT/UVES
spectrum of CD\,--38$^{\circ}$245, the previously most metal-poor
star, and with HE\,0107--5240. The spectrum of
CD\,--38$^{\circ}$245 has been obtained with the same observational
setup. The spectra are on the same scale and have been offset
arbitrarily in the vertical direction. Note the very weak or absent
Fe lines in the spectrum of HE\,0107--5240, and the presence of the
C$_2$ band at $\sim 5165$\,{\AA}.}
\vfill
\pagebreak

\subsection{HE~0107--5240}

Christlieb et\,al.\ (2002a) recently announced the discovery of
HE\,0107--5240, a giant with $\mbox{[Fe/H]}=-5.3$. This star
provides constraints on low-mass star formation in the early
Galaxy, and insights into the properties of the first generation of
stars and supernovae. The abundance pattern of HE\,0107--5240 is
characterized by huge over-abundances of carbon and nitrogen with
respect to iron ($\mbox{[C/Fe]}=+4.0$\,dex;
$\mbox{[N/Fe]}=+2.3$\,dex), while the abundances of the other five
elements of with presently detected spectral lines exhibit a
behavior similar to what is seen in other EMP stars.

Umeda \& Nomoto (2003) proposed \emph{a posteriori} that the gas
cloud from which HE\,0107--5240 formed could have been enriched by
a $25\,M_{\odot}$ Population~III star exploding as a supernova of
low explosion energy ($E_{\mbox{\scriptsize exp}}=3\cdot
10^{50}\,\mbox{erg}$). By assuming that the material produced
during the SN event is homogeneously mixed over a wide range of the
mass coordinate, and that a large fraction of the material falls
back onto the compact remnant (the ``mixing and fallback''
mechanism), Umeda \& Nomoto are able to reproduce the abundance
pattern of HE\,0107--5240 remarkably well (see their Figure 1). In
particular, very high C/Fe and N/Fe ratios can be produced in this
scenario (with the CNO elements produced in late stages of the
evolution of the SN progenitor), offering an interesting
possibility for explaining why a large fraction of the extremely
metal-poor stars exhibit high enhancements of carbon (for a similar
work on the abundance patterns of CS~29498-043 and CS~22949-037 see
Tsujimoto \& Shigeyama 2003).

The scenario of Umeda \& Nomoto is also very attractive in that it
does not require the existence of a binary companion for the CEMP
stars. In fact, there are at least three unevolved CEMP stars known
that do not show any radial velocity variations larger than
0.4~km\,s$^{-1}$ over a period of 8 years (Preston \& Sneden 2001),
ruling out, at least for these three stars, the scenario in which a
formerly more massive star went through its AGB phase and
transfered dredged-up material onto the surface of the less massive
companion, which we observe as a CEMP star today.

Umeda \& Nomoto predict an O abundance of $\mbox{[O/Fe]}=+2.8$\,dex
for HE\,0107--5240. A high-resolution, high $S/N$ spectrum of
HE\,0107--5240, covering the near UV region down to the atmospheric
cutoff, has now been obtained with VLT/UVES. Spectrum synthesis
calculations suggest that the detection of UV-OH lines is possible
if $\mbox{[O/Fe]}>+0.3$\,dex. Therefore, the predictions of Umeda
\& Nomoto can soon be tested.

\subsection*{Acknowledgements}

I am grateful to D. Reimers (P.\,I.\ of the HES) and L. Wisotzki
(HES project scientist) for making my work on the HES possible. It
is a pleasure to acknowledge the contributions of my numerous
collaborators to HES metal-poor star research over the last years.
Here I would like to especially thank those who have participated
in the huge effort of vetting HES candidates: Paul Barklem, Tim
Beers, Mike Bessell, Innocenza Busa, Judy Cohen, Doug Duncan, Bengt
Edvardson, Lisa Elliot, Birgit Fuhrmeister, Dionne James, Andreas
Korn, Andy McWilliam, Solange Ram{\'i}rez, Johannes Reetz, Jaehyon
Rhee, Silvia Rossi, Sean Ryan, Steve Shectman, Ian Thompson, and
Franz-Josef Zickgraf. I thank Tim Beers for valuable comments, and
proof-reading. This work is supported by Deutsche
Forschungsgemeinschaft under grant Re~353/44-1, and was partly
carried out during a Marie Curie Fellowship of the European
Commission (contract number HPMF-CT-2002-01437) at Uppsala
Astronomical Observatory, and a Linkage International Fellowship of
the Australian Research Council (Project ID LX02114310) at Mt.
Stromlo Observatory.

\subsection*{References}

\def\bref{\vspace{3pt}\noindent\hangindent=10mm}

{\small

\bref
Aoki, W., Norris, J.E., Ryan, S.G., Beers, T.C., Ando, H.
2000, ApJ 536, L97

\bref
Aoki, W., Norris, J.E., Ryan, S.G., Beers, T.C., Ando, H.
2002a, ApJ 567, 1166

\bref
Aoki, W., Ryan, S.G., Norris, J.E., Beers, T.C., Ando, H.,
Tsangarides, S.
2002b, ApJ 580, 1149

\bref
Beers, T.C. 1999, in The Third Stromlo Symposium: The Galactic
Halo, ed. B. Gibson, T. Axelrod, \& M. Putman, ASP Conf. Ser. 165,
202--212

\bref
Beers, T.C. 2000, in The First Stars, Proceedings of the second
MPA/ESO workshop, ed. A. Weiss, T. Abel, \& V. Hill (Heidelberg:
Springer), astro-ph/9911171

\bref
Beers, T.C., Preston, G.W., Shectman, S.A. 1985, AJ 90, 2089

\bref
Beers, T.C., Preston, G.W., Shectman, S.A. 1992, AJ 1987

\bref
Beers, T.C., Rossi, S., Norris, J.E., Ryan, S. G., Shefler, T.
1999, AJ 117, 981

\bref
Bonifacio, P., Molaro, P., Beers, T., Vladilo, G. 1998, A\&A 332,
672

\bref
Carretta, E., Gratton, R., Cohen, J., Beers, T., Christlieb, N.
2002, AJ 124, 481

\bref
Cayrel, R. 1996, A\&A  Rev., 7, 217

\bref
Cayrel, R., Hill, V., Beers, T., Barbuy, B., Spite, M., Spite, F.,
Plez, B., Andersen, J., Bonifacio, P., Francois, P., Molaro, P.,
Nordstr\"om, B., Primas, F. 2001, Nature 409, 691

\bref
Christlieb, N., Beers, T. C. 2000, in Subaru HDS Workshop on stars
and galaxies: Decipherment of cosmic history with spectroscopy, ed.
M. Takada-Hidai \& H. Ando (Tokyo: National Astronomical
Observatory), 255--273, astro-ph/0001378

\bref
Christlieb, N., Green, P., Wisotzki, L., Reimers, D. 2001a, A\&A
375, 366

\bref
Christlieb, N., Wisotzki, L., Reimers, D., Homeier, D., Koester,
D., Heber, U. 2001b, A\&A 366, 898

\bref
Christlieb, N., Bessell, M., Beers, T., Gustafsson, B., Korn, A.,
Barklem, P., Karlsson, T., Mizuno-Wiedner, M., Rossi, S. 2002a,
Nature 419, 904

\bref
Christlieb, N., Wisotzki, L., Gra\ss hoff, G. 2002b, A\&A 391, 397

\bref
Cohen, J., Christlieb, N., Beers, T., Gratton, R., Carretta, E.
2002, AJ 124, 26

\bref
Cohen, J., Christlieb, N., Qian, Y., Wasserburg, G. 2003, ApJ, in
press, astro-ph/0301460

\bref
Cowan, J.J., McWilliam, A., Sneden, C., Burris, D.L. 1997, ApJ 480,
246

\bref
Cowan, J.J., Pfeiffer, B., Kratz, K.-L., Thielemann, F.-K., Sneden,
C., Burles, S., Tytler, D., Beers, T. 1999, ApJ 521, 194

\bref
Fuhrmann, K. 1998, A\&A  338, 161

\bref
Goriely, S., Siess, L. 2001, A\&A 378, L25

\bref
Hill, V., Barbuy, B., Spite, M., Spite, F., Plez, R.C.B., Beers,
T., Nordstr\"om, B., Nissen, P. 2000, A\&A 353, 557

\bref
Hill, V., Plez, B., Cayrel, R., Nordstr\"om, T.B.B., Andersen, J.,
Spite, M., Spite, F., Barbuy, B., Bonifacio, P., Depagne, E.,
Fran\c cois, P., Primas, F. 2002, A\&A 387, 560

\bref
Karlsson, T., Gustafsson, B. 2001, A\&A 379, 461

\bref
Lucatello, S., Gratton, R., Carretta, E., Cohen, J., Christlieb, N.,
Beers, T., Ram\'{\i}rez, S. 2003, AJ 125, 875

\bref
McWilliam, A., Preston, G., Sneden, C., Searle, L. 1995, AJ 109,
2757

\bref
Norris, J.E., Ryan, S.G., Beers, T.C. 1997a, ApJ 488, 350

\bref
Norris, J.E., Ryan, S.G., Beers, T.C. 1997b, ApJ 489, L169

\bref
Norris, J., Ryan, S., Beers, T. 2001, ApJ 561, 1034

\bref
Norris, J., Ryan, S., Beers, T., Aoki, W., Ando, H. 2002, ApJ 569,
L107

\bref
Preston, G. 2000, PASP 112, 141

\bref
Preston, G., Sneden, C. 2001, AJ 122, 1545

\bref
Rhee, J. 2000, PhD thesis, Michigan State University

\bref
Rossi, S., Beers, T.C., Sneden, C. 1999, in The Third Stromlo
Symposium: The Galactic Halo, ed. B. Gibson, T. Axelrod,
\& M. Putman, ASP Conf. Ser. 165, 264--268)

\bref
Ryan, S., Norris, J., Beers, T. 1996, ApJ 471, 254

\bref
Ryan, S.G., Norris, J.E., Beers, T.C. 1999, ApJ 523, 654

\bref
Schatz, H., Toenjes, R., Pfeiffer, B., Beers, T., Cowan, J., Hill,
V., Kratz, K. 2002, ApJ 579, 626

\bref
Shigeyama, T., Tsujimoto, T. 1998, ApJ 507, L135

\bref
Skrutskie, M., Schneider, S., Stiening, R., Strom, S., Weinberg,
M., Beichman, C., Chester, T., Cutri, R., Lonsdale, C., Elias, J.,
Elston, R., Capps, R., Carpenter, J., Huchra, J., Liebert, J.,
Monet, D., Price, S., Seitzer, P. 1997, in The Impact of Large
Scale Near-IR Sky Surveys, ed. F. Garz\'on, N. Epchtein, A. Omont,
W. Burton, \& P. Persi (Dordrecht: Kluwer), 25--32

\bref
Sneden, C., McWilliam, A., Preston, G.W., Cowan, J.J., Burris,
D.L., Amorsky, B.J. 1996, ApJ 467, 819

\bref
Sneden, C., Cowan, J., Ivans, I., Fuller, G., Burles, S., Beers,
T., Lawler, J. 2000, ApJ 533, L139

\bref
Tsujimoto, T., Shigeyama, T., Yoshii, Y. 2000, ApJ 531, L33

\bref
Tsujimoto, T., Shigeyama, T. 2003, ApJ 584, L87

\bref
Umeda, H., Nomoto, K. 2003, Nature, in press, astro-ph/0301315

\bref
Van Eck, S., Goriely, S., Jorissen, A., Plez, B. 2001, Nature 412,
793

\bref
Wanajo, S., Itoh, N., Ishimaru, Y., Nozawa, S., Beers, T. 2002, ApJ
577, 853

\bref
Watson, F. 1998, AAO Newsletter, 11

\bref
Wisotzki, L., Christlieb, N., Bade, N., Beckmann, V., K\"ohler, T.,
Vanelle, C., Reimers, D. 2000, A\&A 358, 77

}

\def\bref{\vspace{4pt}\noindent\hangindent=10mm}

\vfill
\pagebreak
\end{document}